\documentclass[aps,pre,print,groupedaddress,amsmath,amssymb]{revtex4-1}

\usepackage{graphics}
\def\rRichClub{\Phi}
\def\kRichClub{\phi}

\def\text#1{\rm{#1}}
\def\densityRC{E}
\def\linksRC{\Delta E}
\def\probLinking{T}
\def \probRankRank{P}
\begin{document}

\title{Random Networks with given rich--club coefficient}
\author{Ra\'ul~J.~Mondrag\'on}
\email[]{r.j.mondragon@eecs.qmul.ac.uk}
\affiliation{Queen Mary University of London, School of Electronic Engineering and Computer Science, Mile End Road, London, E1 4NS, UK}

\author{Shi~Zhou}
\email[]{S.Zhou@cs.ucl.ac.uk}
\affiliation{University College London, Dept. of Computer Science, Gower Street, London WC1E 6BT, UK}

\date{\today}

\begin{abstract}
In complex networks it is common to model a network or generate a surrogate network based on the conservation of the network's degree distribution. We provide an alternative network model based on the conservation of connection density within a set of nodes. This density is measure by the rich--club coefficient. We present a method  to generate surrogates networks with a given rich--club coefficient. We show that by choosing a suitable local linking term, the generated random networks can reproduce the degree distribution and the mixing pattern of real networks. The method is easy to implement and produces good models of real networks. 
\end{abstract}

\pacs{89.75.Fb}
\maketitle

\section{Introduction}
Many complex networks are single networks in the sense that their structure is unique not one of several. In general, we do not have the equivalent of a physical law to verify if the  statistical measures obtained from a single network are expected or exceptional. Instead, a common technique to analyze the properties of a single network is to use  statistical randomization 
methods~\cite{Manly97} to create a reference network which is used for comparison purposes. The procedure consists of using the observed network to generate an ensemble of networks via randomization, usually the reshuffling of the links.  The reference network
is generated from this ensemble and it is used to assess the significance of a property of the network.  To take into consideration the intrinsic structure of the network it is common to use a restricted randomization procedure. The restriction is to reshuffle the connections of the nodes without changing the degree distribution of the nodes $P(k)$~\cite{Molloy98}, where $P(k)$ is the degree distribution, that is, the fraction of nodes in the network with degree $k$.

Many complex networks contain a small set of nodes that have large numbers of links, the so--called rich nodes. In some  networks the rich nodes tend to be tightly interconnected between themselves, forming a rich--club~\cite{Zhou2004}. The rich--club is an oligarchy in that it dominates the organization of the whole network. In scale--free networks~\cite{Albert02} the connectivity of the rich--club plays an important role in the functionality of the network, for example in the transmission of rumors in social networks~\cite{Masuda06} or the efficient delivery of information in the Internet~\cite{Zhou04G}. The rich--club coefficient measures the density of connection among these group of nodes. This set of nodes can be defined by a ranking scheme~\cite{Zhou2004}, by their degree~\cite{Colizza06} or by  a network hierarchy~\cite{McACosCae07}. If the nodes are ranked by non--increasing order of their degree, i.e.~the best-connected node is ranked as $r=1$, the second best-connected node is $r=2$ and so on,  then the {density} of connections between the first $r$ nodes is quantified by the rich--club coefficient~\cite{Zhou2004}
\begin{equation}\label{eq:richRank}
\rRichClub(r)=\frac{2\densityRC(r) }{r(r-1)},
\end{equation}
where $\densityRC(r)$ is the number of links between the $r$ nodes and $r(r-1)/2$ is the maximum number of links that these nodes can share. As a function of the nodeÕs degree, the rich--club coefficient is~\cite{Colizza06}
\begin{equation}
\label{eq:richDegree}
\kRichClub(k)=\frac{2E_{ k}}{N_{ k}(N_{ k} - 1)},
\end{equation}
where $N_k$ is the number of nodes with degree equal or higher than $k$ and $E_k$ is the number of links among these $N_k$ nodes. It is known that the rich--club coefficient $\kRichClub(k)$ is a projection of the degree--degree correlation~\cite{Colizza06}, is independent of its degree distribution~\cite{Zhou07a} and it is non-trivially related to other properties like the clustering coefficient~\cite{Xu10}.

It is possible to construct a surrogate network where the degree distribution $P(k)$ is conserved~\cite{Molloy98,Maslov02}, the rich--club coefficient $\kRichClub(k)$ is conserved or both of these properties are conserved~\cite{Zhou07a}.  Our aim here, is to build a network model and surrogate random networks based on the rank rich--club coefficient $\rRichClub(r)$ and show that they are good models of scale--free networks as they tend to reproduce key statistical properties.

Before introducing the model, we would like to remark that the shape of the function $\rRichClub(r)$ depends on the way that the nodes are ranked. There is an ambiguity when labeling the nodes via a degree--dependent rank. For high degree nodes this is not a problem as the degree tend to be unique so the rank labels these nodes unambiguously. For lower degree nodes, there are many nodes with the same degree. In this case the labeling of the nodes is not unique. It is possible to reduce greatly this redundancy of the labels by using a linear order relationship~\cite{Nagle66}, where the degree of the second neighbors is used to disambiguate the labeling. This linear ordering has been used to visualize higher order  correlations in the network topology~\cite{Guo07,Mondragon08}. We observed that using a degree--only ranking scheme or a linear order ranking has little effect on the statistical properties under consideration here. So in this communication we consider only a degree--dependent ranking scheme.

\begin{figure}
\begin{center}
\includegraphics{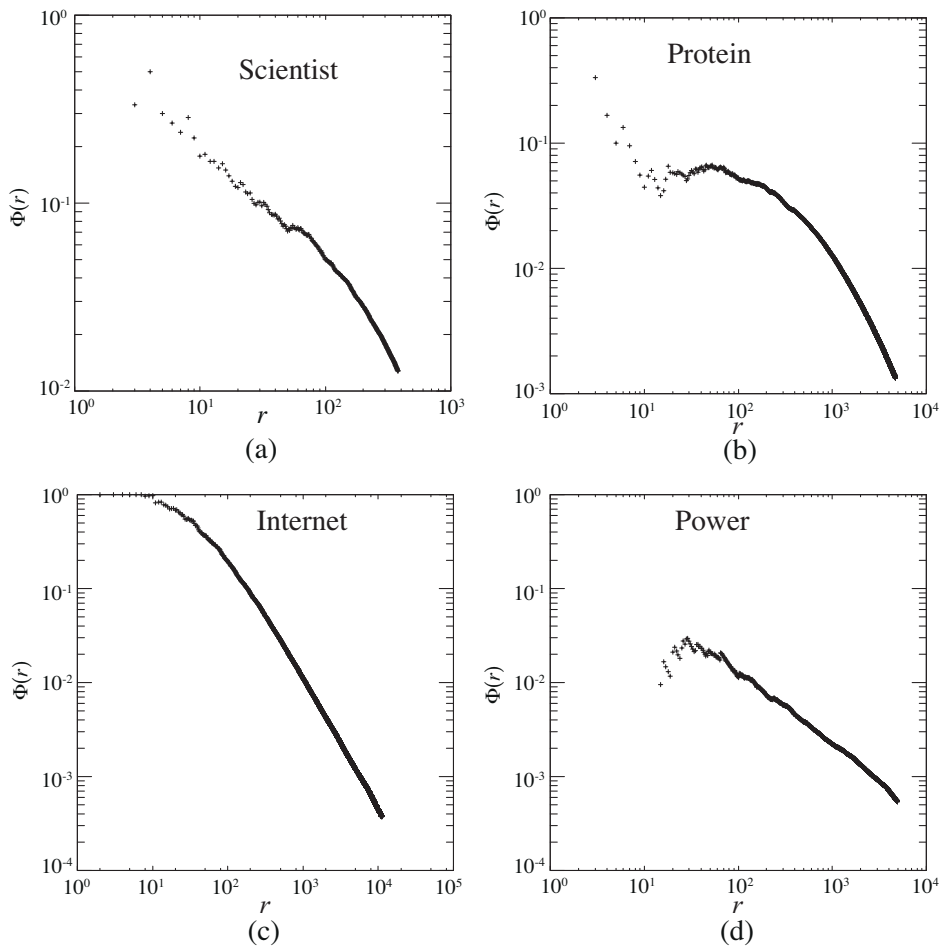}
\end{center}
\caption{\label{fig:one}The rich--club coefficient $\rRichClub(r)$ for the four networks under study.}
\end{figure}

Figure~\ref{fig:one} shows $\rRichClub(r)$ for the Scientists co--authorship (Scientists)~\cite{Newman01}, the protein interaction  (Protein)~\cite{Colizza05}, the AS--Internet (Internet)~\cite{Mahadevan06} and the power grid networks (Power)~\cite{Watts98}. We selected these networks as they have been widely used  when studying the topological properties of complex network as they have different topological characteristics (e.g. \cite{jam09}). These networks have very different connectivities between the high degree nodes. The Internet data has a fully connected core where the top seven nodes are  fully connected ($\rRichClub(7)= 1$) in contrast with  the power grid network where the top fourteen nodes do not share a link at all ($\rRichClub(14)=0$).

\section{Networks defined by the rich--club coefficient}
From equation~(\ref{eq:richRank}) we have that the number of links that node $r$ shares  with the $r-1$ nodes of higher or equal degree is 
\begin{equation}
\label{eq:distdegrees} 
\linksRC(r) = \densityRC(r)-\densityRC(r-1)=\rRichClub(r)\frac{r(r-1)}{2} -
\rRichClub(r-1)\frac{(r-1)(r-2)}{2}
\end{equation}
where $\linksRC(1)=0$, $\sum_{i=1}^N \linksRC(i)=L$ where $L$ is the total number of links and $N$ is the total number of nodes. The term $\linksRC(1)$ is zero,  because  the node $r=1$  is the top node. The number of links $\linksRC(r)$ is bounded by $0 \le \linksRC(r) \le r-1$. If $\linksRC(r)$ is known then for $r \ge 2$,  $\rRichClub(r)$ is obtained from the recursive equation
\begin{equation}
\label{eq:recursive}
\rRichClub(r) = \frac{2\linksRC(r) + (r-1)(r-2)\rRichClub(r-1)}{r(r-1)}
\end{equation}
with $\rRichClub(1) = 0$.

To determine the network connectivity we need to distribute  $\linksRC(r)$ links between node $r$ and the  $r'\in [1, r-1]$ nodes, for all $r$. Let us assume that $P(r',r)$ is the probability that node $r$ connects to node $r'<r$, $\probRankRank(r,r)=0$ as self--loops are not allowed and two nodes can share only one link. Given  $\linksRC(r)$ links for $1\le r \le N$, we  constrain the connectivity of a network by imposing the condition that the average number of links, $\overline{\linksRC}(r)$, between  node $r$ and the $r-1$ nodes of lower rank  is
\begin{equation}
\label{eq:probRestriction}
\overline{\linksRC}(r) = \sum_{i=1}^{r-1} P(i,r)=\linksRC(r).
\end{equation}
This last equation defines an ensemble of networks where their average rich--club coefficient $\bar\rRichClub(r)$ is defined by equations~(\ref{eq:recursive}) and (\ref{eq:probRestriction}). A network  from this ensemble satisfies that 
the average degree $\bar k_r$ of node $r$ is
\begin{equation}
\label{eq:estimation-k} 
\bar k_r =\sum_{r'=1}^N\probRankRank(r',r)=\linksRC(r) + \sum_{j=r+1}^N \probRankRank(r,j)
\end{equation} 
with standard deviation $\sigma_{k_{r}}^2=\sum_{r'=1}^N\probRankRank(r',r)(1-\probRankRank(r',r))$.
The average number of links is  $\bar L=\sum_{r}\sum_{r'} \probRankRank(r',r) $ with standard deviation $\sigma_L^2=\sum_{r}\sum_{r'} \probRankRank(r',r)(1-\probRankRank(r',r))$
and  the average degree of the nearest--neighbors~\cite{Pastor01} of a node with degree $k$ is 
\begin{equation}
\label{eq:estimation-knn}
k_{\text{nn}}(k) = \sum_{k'=1}^{k_{\rm{max}}} k' P(k'|k) \approx \frac{1}{N_k}\sum_{i=1}^{N}\left(\frac{1}{k}
\sum_{j=1}^{N} \probRankRank(i,j) k_j\right)\delta_{k_i,k}.
\end{equation}
In the above equation, the Kronecker delta is introduce to  consider only  nodes with degree equal to $k$,  $N_k$ is the number of $k$--degree nodes and the term $1/k$ is a normalization factor.

\subsection{Local linking term $\probLinking(r',r)$}
As an ansatz we assume that the probability that there is a link between node $r$ and $r'$ can be factorized as
\begin{equation}
\label{eq:probRankRank}
P(r', r) =\probLinking(r',r)\,\linksRC(r), \quad r'<r,
\end{equation} 
where $\probLinking(r',r)$ is a local linking factor. From equation~(\ref{eq:probRestriction}) we have that $\sum_{r'=1}^{r-1} T(r',r)=1$.  
In here the sample space is the set of all possible combination of links that the node has with the $r-1$ nodes of lower rank, with the restriction that only one link is shared between two nodes. The set of events, used in the definition of the probability, is the different combination that $\linksRC(r)$ links can be shared between node $r$ and the $r-1$ lower rank nodes. Notice that  $\linksRC(r)$ is bound by $0\le \linksRC(r) \le r-1$. 

\subsubsection{Egalitarian linking} 
The simplest case is when the $\linksRC(r)$ links that node $r$ can share with the $r-1$ nodes are evenly distributed,  then the probability that node $r$ connects to $r'$ is
\begin{equation}
\label{eq:egalitarian}
P(r',r) = \frac{1}{r-1} \linksRC(r), \quad r'<r,
\end{equation}
where $\probLinking(r',r) =1/(r-1)$. For example if node $r$ has a link with all the $r-1$ nodes of lower rank, i.e. $\linksRC(r)=r-1$ then $P(r',r)=1$, $r'<r$.

\subsubsection{Preferential linking}
For the case that node $r$ prefers to connect to nodes with lower rank (i.e. higher degree), we propose  the preferential linking term $\probLinking(r',r)=r'^{-\alpha}/S(r)$, where $\alpha>0$ is a constant and $S(r)$ is a normalization factor. The probability that  there is a link between $r$ and $r'$ is 
\begin{equation}
\label{eq:preferential}
P(r', r) = \frac{r'^{-\alpha}}{S(r)}\linksRC(r)=\left( \frac{r'^{-\alpha}}{\sum_{i=1}^{r-1} i^{-\alpha} }\right)\linksRC(r),
\end{equation}
where $S(r)=\sum_{i=1}^{r-1} i^{-\alpha}$ to ensure that $\sum_{i=1}^{r-1} P(i,r)=\linksRC(r)$ (see equation~\ref{eq:probRestriction}).
Note that it is possible that for some networks the probability function given by equation~(\ref{eq:preferential}) may be larger than 1, in that case a more suitable $\probLinking(r',r)$ should be considered.

\subsection{Evaluation of the model}
\begin{figure}
\begin{center}
\includegraphics{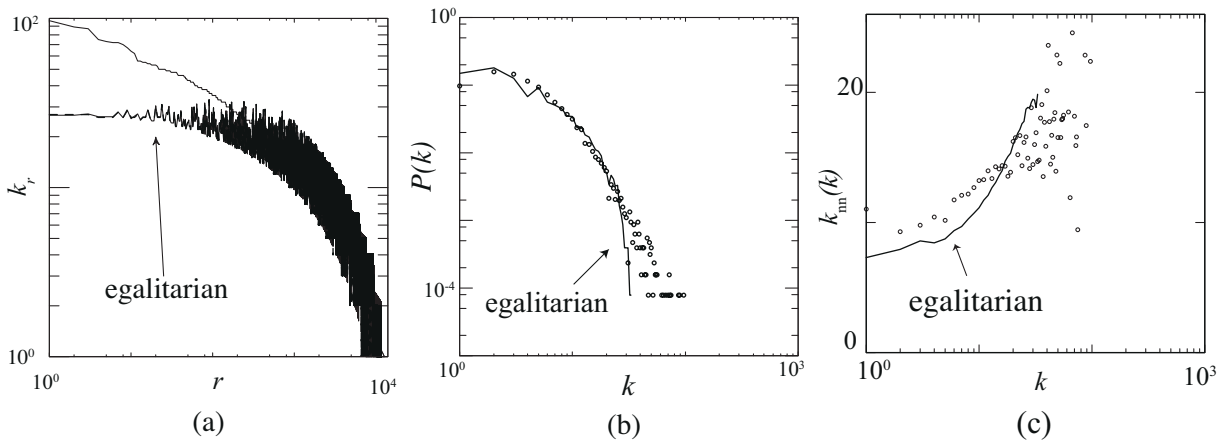}
\end{center}
\caption{\label{fig:two}Topological properties of an egalitarian rich--club network constructed from the Scientists co--authorship network.}
\end{figure}

\begin{figure}
\includegraphics{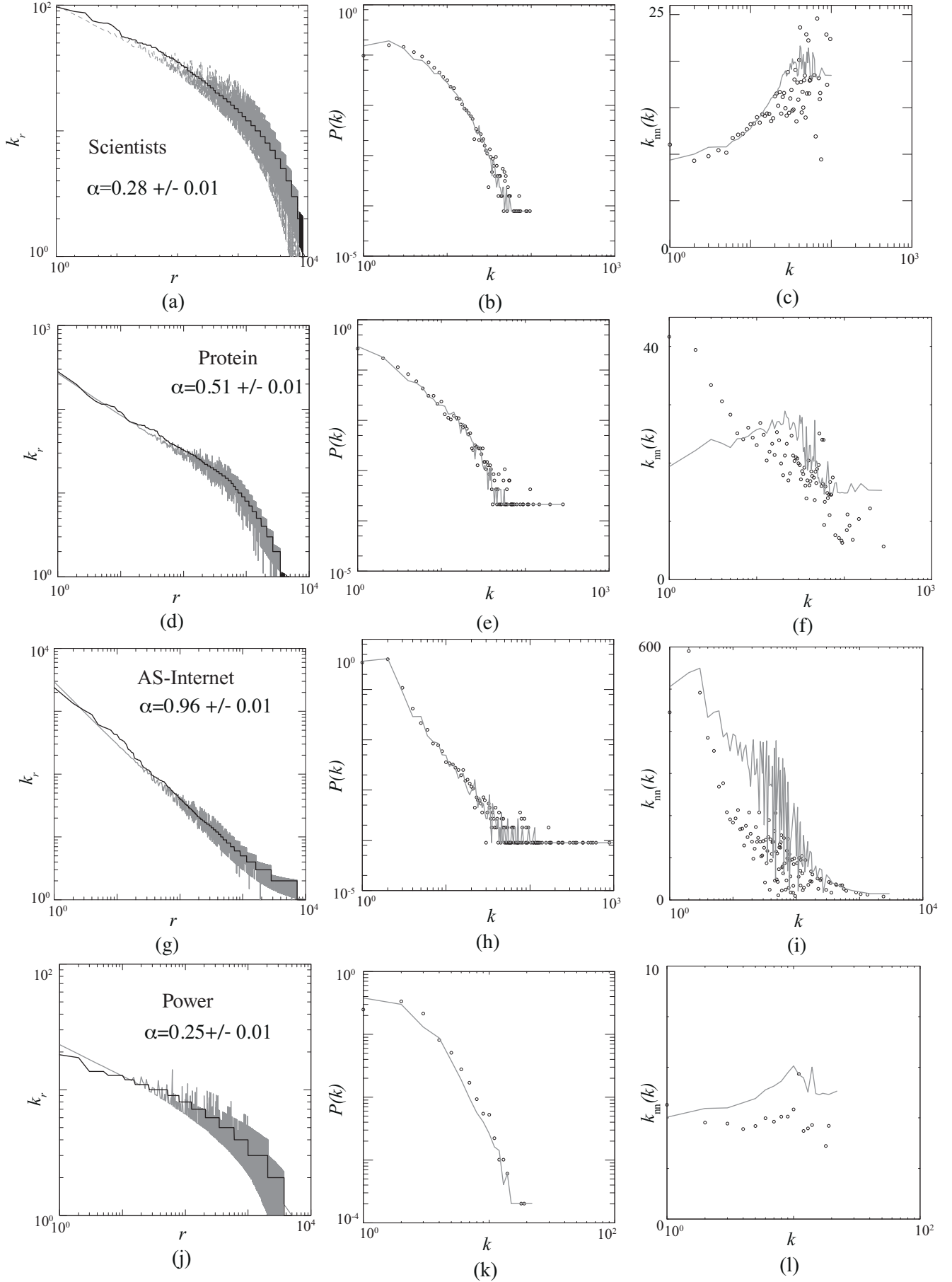}
\caption{The node's degrees $k_r$, their degree distribution $P(k)$ and average degree of the nearest--neighbors $k_{\text{nn}}$ for the four networks under study. The properties of the original network are show in black and  their models in gray. 
For these $\alpha$ values, using linear regression of $(k_r-\bar k_r )$ vs. $r$, the regression coefficient $R$ of the networks is between $0.85$ and $0.87$.}
\label{fig:three}
\end{figure}

Figure~\ref{fig:one}(a) shows the rich--club coefficient  $\rRichClub(r) $ as a function of node rank $r$ for the Scientists network. From the data we evaluated $\rRichClub(r)$ and $\linksRC(r)$. To build a model of the Scientists network we assume that the local linking term is egalitarian  and   evaluated $ \probRankRank(r,r')$,  $\bar k_r$, and $k_{\text{nn}}(k)$ using equations\,(\ref{eq:egalitarian}), (\ref{eq:estimation-k}) and (\ref{eq:estimation-knn}). To obtain an integer value of the node's degree we used the integer function $| \bar k_r |$. The degree distribution $P(k)$ was obtained using $| \bar k_r |$ for all $r$.  Figures~\ref{fig:two}(a) to (c) show that our model resembles  the real network, except that the top ranked nodes have similar degrees, which is  not true in the real network  (see $r<20$ in Figure~\ref{fig:two}(a)).   In fact these nodes have significantly different degrees, which suggests that there is a preferential linking between the high degree nodes.  To verify this last statement we created a network model using the preferential linking defined in equation~(\ref{eq:preferential}). The  exponent $\alpha$ was obtained by fitting the average degree of node $\bar k_r$  using equation~(\ref{eq:estimation-k}) against the degree $k_r$ from the original network, we did so by obtaining the value of $\alpha$ that minimizes  the square of the error $\sum_r(k_r-\bar k_r )^2$ to a precision of  $10^{-2}$ in $\alpha$.
Figures~\ref{fig:three}(a)--(c) show that a model based on the preferential linking is a good approximation of the Scientists network. 

We also created a  network model based on preferential linking for the Protein interaction network~\cite{Colizza05},  the AS--Internet~\cite{Mahadevan06} and  the Power grid network~\cite{Watts98}.  Figures~\ref{fig:three}(d)--(l) demonstrate that our models closely resemble the degree distribution and give a good approximation to the nearest-neighbors average degree for both assortative (co--authorship) and disassortative (Internet, Protein and Power grid) networks.

\section{Generating Random Networks With Given Rich-Club Coefficient}
For a given network we can generate a surrogate random network which preserves  the total number of nodes $N$, links $L$, and  the rich--club coefficient 
$\rRichClub(r)$ for all $r$.  First the $N$ nodes are ranked by their degree and the number of links $\linksRC(r)$ is evaluated from $\rRichClub(r)$. To create a surrogate network, we create $N$ stub nodes by assigning to each  node $r\in[2,N]$, $\linksRC(r)$ links. From $r=2$ to $N$, we connect the 
$\linksRC(r)$ links between the stub node $r$ and the smaller--ranked $r'<r$  nodes. The probability that node $r$ connects to node $r'$ is defined by equation~(\ref{eq:probRankRank}).
 We do not allow self--links or that two nodes share more than one link. Notice that $\sum_{r=1}^N \linksRC(r) = L$.
All of these surrogate networks will have the same rich--club coefficient as the original network because all of them have the same  $\linksRC(r)$ for all $r$. 

\begin{figure}
\includegraphics{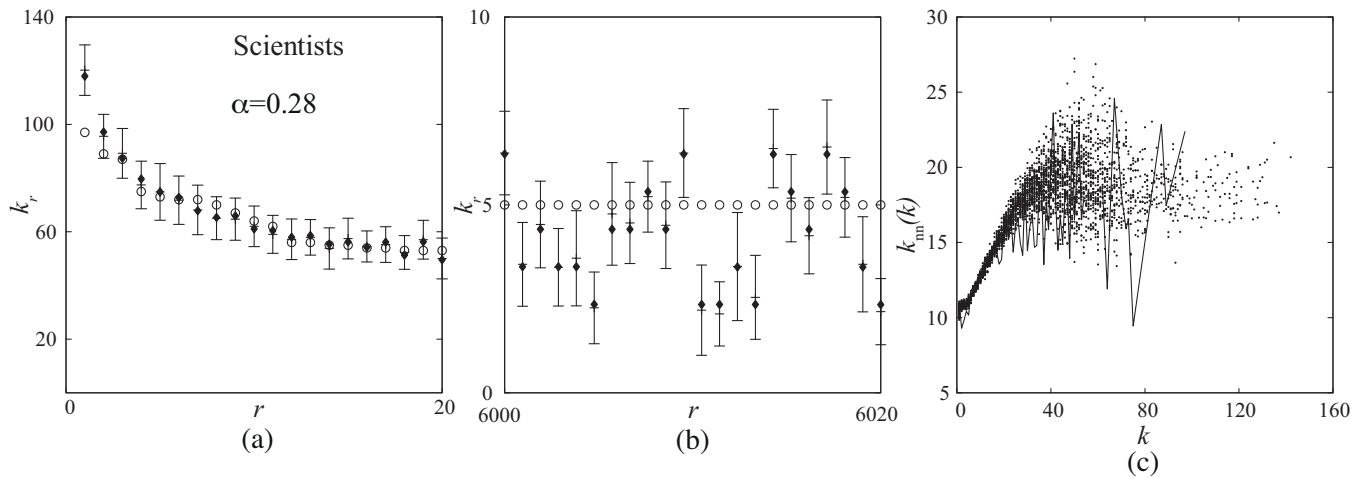}
\caption{(a) Degree $k_r$ of the 20 smallest ranked nodes (i.e. $1\le r\le 20$), and (b) for $6000\le r \le 6020$. The open circle is the original data, the filled diamonds is the average obtained from equation~(\ref{eq:estimation-k}) and the line with the error bars obtained from 40 realizations of the random networks. (c) The nearest--neighbors average degree $k_{\text{nn}}(k)$ for the 40 random networks (dots) compared against the value of the original network (solid line). }
\label{fig:molRed}
\end{figure}

As an example, we illustrate the technique using the Scientists co--authorship collaboration network using preferential linking. Figure~\ref{fig:molRed}(a)--(b) show the average node degree as a function of node  rank, obtained from 40 surrogate random networks generated using the  preferential linking with $\alpha = 0.28$. The average degree of the random networks closely (lines with error bars)  approximates the degree of the original network (open circles). 
In the figure we also plotted the average degree obtained from equation~(\ref{eq:estimation-k}) (filled diamonds). Notice that we expect a discrepancy between the average degree evaluated from the random networks and the average degree obtained from equation~(\ref{eq:estimation-k}). The reason is that the restriction imposed by equation~(\ref{eq:probRestriction}) means that the \emph{average} number of links between node $r$ and the $r'<r$ nodes is $\linksRC(r)$. However, the random surrogate networks are generated with the restriction that the number of links between node $r$ and the $r'<r$ nodes is exactly $\linksRC(r)$.
 
Figure~\ref{fig:molRed}(c) compares the nearest--neighbors average degree  $k_{\text{nn}}(k)$ of the original network (line) against the value obtained by the 40  random networks (scatter dots). It is clear that the random networks resemble well the assortative property of the real network.  

\section{Conclusion}
We investigated the properties of random networks defined by a rank--based rich-club coefficient.  We show that using a preferential local linking term, we can approximate the degree distribution and the mixing pattern of a variety of real networks. We also introduced a method to generate surrogate random networks  conserving the network's rich--club coefficient. While existing surrogate network  models have focused on connectivity of individual nodes, our work provides an alternative  method based on the hierarchical structure of a network in terms of link density  among a group of nodes. 

\begin{acknowledgments}
We thank the referees and Rui Carvalho for their useful comments and suggestions. SZ is supported by the Royal Academy of Engineering. Both authors thanks the UK Engineering and Physical Sciences Research Council (EPSRC) for support under grant number 10216/70 (SZ) and EP/H04812X (RJM).
\end{acknowledgments}

\bibliography{RichClubNull}

\end{document}